\documentclass[a4paper,11pt]{article}
\usepackage{jheppub} % for details on the use of the package, please see the JINST-author-manual
\usepackage{lineno}
\usepackage[english]{babel}
\usepackage{simplewick}
\usepackage{verbatim}
\usepackage{epsfig}
\usepackage{latexsym}
\usepackage{amsmath}
\usepackage{amssymb}
\usepackage{amsfonts}
\usepackage{graphicx}
\usepackage{appendix}
\usepackage{multirow}
\usepackage{blindtext, rotating}
\usepackage[pdftex]{pict2e}
\usepackage{slashed}
\usepackage{physics}
\usepackage{leftidx}

\usepackage{hyperref}
\usepackage{cleveref}
\usepackage{comment}

\newcommand{\be}{\begin{equation}}
\newcommand{\ee}{\end{equation}}
\newcommand{\bea}{\begin{eqnarray}}
\newcommand{\eea}{\end{eqnarray}}
\newcommand{\id}{1\!\!1}

\newcommand{\lb}{\left\lbrace}
\newcommand{\rb}{\right\rbrace}
\newcommand*\diff{\mathop{}\!\mathrm{d}}
\newcommand{\rmii}[1]{{\mbox{\tiny\rm{#1}}}}

\author[a,b]{Marco Cè}
\author[a,b]{Leonardo Giusti}
\author[c]{Davide Laudicina}
\author[b]{Michele Pepe}
\author[a,b]{Pietro Rescigno}

\affiliation[a]{Department of Physics “Giuseppe Occhialini”, University of Milano-Bicocca,\\Piazza della Scienza 3, I-20126 Milano, Italy}
\affiliation[b]{INFN Milano–Bicocca,\\ Piazza della Scienza 3, I-20126 Milano, Italy}
\affiliation[c]{Fakultät für Physik und Astronomie, Institut für Theoretische Physik II, Ruhr-Universität Bochum,\\44780 Bochum, Germany}

\title{The hyperfine splitting in QCD mesonic screening masses at asymptotically large temperatures}

\emailAdd{marco.ce@unimib.it}
\emailAdd{leonardo.giusti@unimib.it}
\emailAdd{davide.laudicina@ruhr-uni-bochum.de}
\emailAdd{michele.pepe@mib.infn.it}
\emailAdd{p.rescigno1@campus.unimib.it}

\abstract{We determine the hyperfine splitting in the QCD flavour non-singlet mesonic
screening masses at asymptotically large temperatures. The analytic calculation
is carried out in the dimensionally-reduced effective theory where the first non-zero
contribution is of $O(g^4)$ in the QCD coupling constant $g$. Apart for its own theoretical
interest, this result provides instrumental information to interpret and to 
parameterize non-perturbative data that are being produced at very high temperatures
by numerical simulations of lattice QCD.  Indeed, the comparison with existing non-perturbative
results shows that higher order (non-perturbative)
contributions in $g$ are needed to explain the data up to the highest temperatures explored,
which is of the order of the electroweak scale.}

\begin{document}
\maketitle
\flushbottom

\section{Introduction}
The dynamics of the strong interactions at high temperature plays a key r\^ole in several physical processes, from the
cosmological evolution of the early universe,
to the interpretation of the experimental data at the relativistic heavy ion collision facilities. Since Quantum
Chromodynamics (QCD) is asymptotically free,
one could hope that, for sufficiently high temperatures, its dynamics can be efficiently described perturbatively.
Nevertheless, at asymptotically high temperatures, QCD effectively behaves as a three-dimensional Yang-Mills
theory which exhibits confinement and, therefore, has to be solved non-perturbatively~\cite{Linde:1980ts}.
As a consequence, the perturbative series in the strong coupling constant $g$ of a given physical quantity
can only be computed up to a certain order,
after which non-perturbative effects start to appear. A well-known example is
the QCD Equation of State for which it is possible to compute perturbatively only terms up
to $O(g^6\ln(g))$~\cite{Braaten:1995jr,Kajantie:2002wa}. A further issue regarding thermal
perturbation theory is the observed slow convergence
of the series, even at very high temperatures. As a result, the non-perturbative dynamics turns out to be relevant
up to temperatures orders of magnitude larger than the GeV scale. This kind of behaviour has been observed first in the
SU($3$) Yang-Mills theory~\cite{Borsanyi:2012ve,Giusti:2016iqr}, and then only very recently also
in QCD~\cite{DallaBrida:2020gux,DallaBrida:2021ddx,Giusti:2024ohu,Bresciani:2025vxw}.

Mesonic screening masses characterize the behaviour of spatially-separated correlation functions.
They are the inverse of spatial correlation lengths, and describe the response of the quark-gluon plasma when
a meson with a given set of quantum numbers is injected into the system. Flavour non-singlet mesonic screening masses
at zero Matsubara frequency are among the simplest quantities that can be computed in the thermal theory, and therefore
they are an ideal playground for comparing analytic results with non-perturbative data. In perturbation theory
they have been calculated at the next-to-leading order in the dimensionally-reduced effective theory some time
ago~\cite{Laine:2003bd}, while baryonic screening masses with nucleon quantum numbers have been computed at the same
order only very recently \cite{Giusti:2024mwq}. Non-perturbatively, the steady conceptual, algorithmic and
technological progress
in lattice QCD has made it possible numerical computations of mesonic and baryonic screening masses
in a wide range of temperature with permille accuracy~\cite{Bazavov:2019www,DallaBrida:2021ddx,Giusti:2024ohu}.

The aim of this paper is to calculate the leading contribution of order $O(g^4)$ to the 
(hyperfine) splitting between the masses at zero Matsubara frequency in the pseudoscalar and 
in the vector channels 
at asymptotically large temperature. The hyperfine splitting is of particular interest in thermal QCD,
as well as for quarkonia in QCD at zero temperature~\cite{Brambilla:2010cs}, because it is sensitive (a)
to all the three scales entering the effective theory, namely the hard $\pi T$, the soft
$m_{\rm E}$ and the ultrasoft $g^2_{\rm E}$ one, and (b) to non-perturbative contributions
to the potential of the relevant Schr\"odinger equation. The high accuracy of the lattice
data that has been reached recently made it possible to clearly resolve the hyperfine splitting between the
masses in the pseudoscalar and in the vector channel up to temperatures of the order of the electroweak
scale~\cite{DallaBrida:2021ddx}. All these facts make the hyperfine splitting an ideal quantity to test the
effective field theory construction against non-perturbative results, and to scrutinize
the range of applicability of its perturbative treatment. Finally, knowing analytically the leading $O(g^4)$
contribution is instrumental to interpret and to parameterize non-perturbative data. 

This paper is organized as follows. In Section~\ref{sec:preliminaries} we define the
interpolating fields and their
two-point correlation functions in QCD, as well as the concept of screening masses. Section~\ref{sec:eft} is devoted to briefly
introduce the effective field theory framework in which the calculation of the hyperfine splitting is carried out.
Here we also introduce the equations of motion and the expression of the quark propagators at leading and
next-to-leading orders. In Section~\ref{sec:mesonic_eft} the interpolating fields already introduced
in Section~\ref{sec:preliminaries} are re-expressed in the context of the three-dimensional effective theory, and
the two-dimensional Schr\"odinger equations satisfied by the corresponding correlation functions are derived.
The computation of the hyperfine splitting is carried out in Section~\ref{sec:results}, while
in Section~\ref{sec:comparison} we compare our result with the non-perturbative data in Ref.~\cite{DallaBrida:2021ddx}.
Finally we discuss our conclusions in Section~\ref{sec:conc}. Additional
technical details are reported in four appendices.

\section{Preliminaries}
\label{sec:preliminaries}
In this paper we focus on the flavour non-singlet pseudoscalar and vector
screening masses at zero Matsubara frequency related to the interpolating fields
\begin{align}
\label{eq:operators}
{\cal O}^a (x_0,x) = \overline\psi(x_0,x) \Gamma_{\cal O} T^a \psi(x_0,x) \,,
\end{align}
where ${\cal O}=\left\{P,V_2\right\}$, $\Gamma_{{\cal O}}=\left\{\gamma_5,\gamma_2\right\}$,
and we make the definite choice of selecting the component for the vector current
in direction $2$. The $T^a$ are the $a=1,\dots,N_f^2-1$ generators of the SU($N_f$) algebra,
normalized so that $\Tr \left[ T^a T^b \right]= \delta^{ab}/2$, which dictate the
flavour structure of the interpolating fields. The two-point screening correlation
functions considered in this paper are
\begin{align}
\label{eq:2pt}
    {\cal C}_{\cal O} (x_3) = \int_0^{1/T} \diff x_0 \int_{\mathbf{R}} \expval{{\cal O}^a (x_0,x) {{\cal O}^a}(0)} \,, \qquad x =  (\mathbf{R},x_3) \, ,
\end{align}
where no summation over the flavour index $a$ is understood. Here
$\int_{\mathbf{R}} = \int \diff^2 \mathbf{R}$ denotes the integration over the transverse spatial directions.
Since we are assuming mass-degenerate quarks, the two-point correlation functions above are
independent on the flavour structure of the interpolating fields, and therefore we drop the flavour
index $a$ on the l.h.s. of eq.~(\ref{eq:2pt}). The corresponding screening masses probe the
exponential fall-off of the screening
correlators at asymptotically large distances, and are defined as
\begin{align}
    m_{\cal O} = -\lim_{x_3\to \infty} \frac{\diff}{\diff x_3} \ln \left[ {\cal C}_{\cal O}(x_3) \right] \, .
\end{align}
The $O(g^2)$ next-to-leading correction to $m_{\cal O}$ was computed in the three-dimensional
effective theory in Refs.~\cite{Hansson:1991kb,Laine:2003bd}. For three massless quarks, its expression reads
\begin{align}
\label{eq:m_nlo}
    m_{\cal O}^{\rmii{nlo}} = 2\pi T \left( 1+0.032740 \cdot g^2 \right) \,,
\end{align}
where the first term on the r.h.s. is the free-field theory value, while the second one is due to interactions.
Notice that, at $O(g^2)$ the value of the screening mass is independent of the field ${\cal O}^a$ defined
in eq.~\eqref{eq:operators}, i.e. it is spin-independent. This is because the hyperfine splitting, defined as
\begin{align}
\label{eq:DeltaM}
    \Delta m_{\rmii{VP}} = m_{\rmii V}-m_{\rmii P} \, ,
\end{align}
starts at $O(g^4)$, see below.

\section{Effective description}
\label{sec:eft}
In thermal QCD at large temperatures, gauge fields in the zero Matsubara
sector are the relevant degrees of freedom at distances much larger than the extent of the
compact direction.  The contributions from the non-zero Matsubara sectors ($n\neq 0$) are
strongly suppressed due to their heavy ``masses'' that are $\sim 2 n \pi T$ ($n\in \mathbb{Z} $).
Quark modes, due to anti-periodic boundary conditions, are always heavy
with ``masses'' $\sim (2n+1) \pi T $. In this context the physics that takes
place at energies much lower than T, or equivalently that involves distances much larger
than the compact direction, can be described by a three-dimensional effective
SU($3$) Yang-Mills theory~\cite{Ginsparg:1980ef,Appelquist:1981vg,Braaten:1995jr}.

\subsection{Effective action}
The effective theory contains gauge fields
$A_k$, with $k=1,2,3$, living in three spatial dimensions. They are coupled to
a massive scalar field $A_0$, which transforms under the adjoint representation of
the gauge group. The action of the effective theory, called Electrostatic QCD (EQCD), reads
\cite{Ginsparg:1980ef,Appelquist:1981vg}
\begin{align}
\label{eq:EQCD}
    S_\rmii{EQCD} \, = \, \int\! {\rm d}^3 x\,
   \lb \frac{1}{2}
   \Tr \left[ F_{jk}F_{jk} \right] 
  +\Tr\left[\left(D_j A_0\right)\left(D_jA_0\right) \right]
  + m_\rmii{E}^2 \Tr \left[ A_0^2 \right]\rb \, + \, \dots \, ,
\end{align}
where the trace is over the color index\footnote{Throughout the paper we use the symbol $\Tr$ to indicate the trace over the indices of the matrix, i.e. the trace can be over colour and/or spin indices. Analogously the symbol $\id$ is
the identity in colour and/or spin indices.}, and the dots stand
for contributions from higher-dimensional fields \cite{Laine:2016hma}. Here $j,k=1,2,3$ and 
$[ D_j,D_k ]=-i g_\rmii{E} F_{jk}$
with the covariant derivative defined as\footnote{In the effective theory we adopt the usual perturbative normalization
of the gauge field.}
$
 D_j=\partial_j-ig_\rmii{E}A_j
$.  
The low-energy constants $m^2_\rmii{E}$ and $g^2_\rmii{E}$ are the Lagrangian mass squared of the scalar field
$A_0$ and the dimensionful coupling constant of the three-dimensional Yang-Mills theory, respectively. They have
been perturbatively matched to QCD in
Refs.~\cite{Kapusta:1979fh,Laine:2005ai,Ghisoiu:2015uza}. At leading order they read
\begin{align}
    m^2_\rmii{E} \, = \, g^2T^2
    \left(\frac{N_c}{3}+\frac{N_f}{6}\right)
    + O(g^4_{ }T^2)
    \,, \qquad
    g^2_\rmii{E} \, =\, g^2 T 
    + O(g^4 T)
   \, , \label{eq:ge}
\end{align}
where $g$ is the QCD coupling constant, $N_f\,$ is the number of flavours and $N_c$ is the number of colours. 
Therefore, at asymptotically high temperatures, the scalar field with mass $m_{\rmii E}$
represents a heavy mode w.r.t. the gauge field. When studying physics that takes place at
energies much lower than $m_{\rmii E}$, the field $A_0$ can be integrated out,
and the action of the resulting effective theory, usually called Magnetostatic QCD (MQCD), reads
\begin{align}
\label{eq:MQCD}
    S_{\rmii{MQCD}} \, = \, \frac{1}{2} \int\! {\rm d}^3 x\,  \Tr \left[ F_{jk}F_{jk} \right]+\dots \, .
\end{align}
Note that, being a three-dimensional Yang-Mills theory with coupling $g_\rmii{M}=g_\rmii{E}+\dots$, its dynamics is
non-perturbative at all temperatures~\cite{Linde:1980ts} with $g_{\rmii M}$ that sets the scale in the theory already
at the classical level.

As a result, QCD develops three different energy scales namely
$g_{\rmii E}^2$, $m_{\rmii E}$, and $\pi T$ which, for sufficiently high temperatures, satisfy the hierarchy
\begin{align}
\label{eq:scales}
    \frac{g_{\rmii E}^2}{\pi} \ll m_{\rmii E} \ll \pi T \quad \longrightarrow \quad \frac{g^2}{\pi^2} \ll \frac{g}{\pi} \ll 1 \, ,
\end{align}
where on the r.h.s. we used the expressions at leading order for $g_{\rmii E}^2$ and $m_{\rmii E}$ in eq. \eqref{eq:ge}.

\subsection{Non-relativistic fermionic action}
As anticipated, at high temperatures quarks are heavy fields and their dynamics can be effectively described by
a $3$-dimensional non-relativistic QCD (NRQCD)~\cite{Huang:1995tz,Caswell:1985ui,Brambilla:1999xf}. The lightest
fermion modes are those in the Matsubara sectors $n=0,-1$ with a ``mass'' of $\pi T$. By following the
derivation in Appendix A of Ref. \cite{Giusti:2024mwq} supplemented by the leading spin-dependent terms, the
effective action for a single flavour in the $n=0$ Matsubara sector,  which couples to soft and
ultrasoft gauge fields, reads
\begin{align}
\label{eq:NRQCD}
    S_{\rmii{NRQCD}} = i \int \diff ^3 x \,\bigg\{ &\bar\chi \left[M-g_{\rmii E}A_0+D_3-\frac{1}{2\pi T}\left(D_\perp^2+\frac{g_{\rmii E}}{4i} [\sigma_j,\sigma_k] F_{jk} \right)\right] \chi \nonumber\\
    +\,&\bar\phi \left[ M - g_{\rmii E}A_0 - D_3 -\frac{1}{2\pi T} \left( D_\perp^2 + \frac{g_{\rmii E}}{4i} [\sigma_j,\sigma_k] F_{jk} \right) \right] \phi\bigg\}+\dots \, ,
\end{align}
where $j,k=1,2$ and $\chi$ and $\phi$ are three-dimensional two-component spinor fields, see
Ref.~\cite{Laine:2003bd,Brandt:2014uda,Giusti:2024mwq}. In eq.~\eqref{eq:NRQCD} the fields $\chi$
and $\phi$ propagate forward and backward in the third spatial direction, respectively. For the
clarity of the presentation we do not report the corresponding contributions for the fields in the fermionic Matsubara
sector $n=-1$, with backward propagating $\chi$ fields and forward propagating $\phi$ fields
in the third spatial direction respectively. These fields contribute to the correlators of interest
for this work with the very same Wick contractions of those in the $n=0$ sector, and we will account
for their contribution in the prefactors of their expressions. In eq.~\eqref{eq:NRQCD} we have
already made explicit the expressions
of the low energy constants appearing in front of the various terms at the order we are interested, e.g.
the coupling appearing in front of the spin-dependent term coincide at this order with $g_{\rmii E}$
given in eq.~\eqref{eq:ge} and appearing in the covariant derivatives. Finally, the expression of $M$ at $O(g^2)$
reads~\cite{Laine:2003bd}
\begin{align}
\label{eq:M}
    M=\pi T \left(1+\frac{g^2}{8\pi^2}C_F\right)\,, \quad\mbox{where}\quad C_F=\frac{N_c^2-1}{2N_c}\, .
\end{align}
The leading contribution to the hyperfine splitting in the mesonic screening masses is due to the
non-diagonal (in spinor indices) term in eq. \eqref{eq:NRQCD}, and it is due to the exchange of
ultrasoft gluons between two quark propagators, whose dynamics is described by
the Yang--Mills part of the action in eq. \eqref{eq:EQCD}.
In analogy with the power counting rules established in Table 1 of Ref. \cite{Giusti:2024mwq}, by looking
at the EQCD action and by taking into account that the relevant dimensionful scale is $g_{\rmii E}^2$,
it follows that
$F_{jk}=O(g_{\rmii E}^3)$. As a consequence, the non-diagonal term (in spinor indices) in eq.~\eqref{eq:NRQCD}  satisfies the
power counting rule
\begin{align}
\label{eq:SP-term}
    g_{\rmii E} \left[ \sigma_j,\sigma_k \right] F_{jk} = O(g_{\rmii E}^4) \,,
\end{align}
which translates, by using eq. \eqref{eq:ge}, into a term of $O(g^4)$.
By taking into account the discussion reported in Appendix~\ref{App:A}, the non-relativistic effective
action in eq.~\eqref{eq:NRQCD} can be written in a compact way as
\begin{align}
\label{eq:NRQCD2}
    S_{\rmii{NRQCD}} = i \int \diff ^3 x \left[ \bar\chi \left( {\cal D}^+ - g_{\rmii E}\, {\cal K}^+ \right) \chi   +
    \bar\phi \left( {\cal D}^- - g_{\rmii E}\, {\cal K}^- \right) \phi \right]\, ,
\end{align}
where the differential operators ${\cal D}^\pm$ and the
interaction vertices ${\cal K}^\pm$ are defined in eqs. \eqref{eq:D} and \eqref{eq:Vpm} respectively.

\subsection{Fermionic equations of motion}
Starting from the action in eq.~\eqref{eq:NRQCD2}, under infinitesimal transformations of the quark fields, we see that,
for each flavour, the $\chi$ and $\phi$ fields satisfy the equations of motion
\begin{align}
i\, \Big\langle\left[{\cal D}^+ - g_{\rmii E}\, {\cal K}^+ \right] \chi(y)\, O(z)\Big\rangle &=\expval{\frac{\delta O(z)}{\delta \bar{\chi} (y)}} \label{eq:EOMc}\\[0.125cm]
i\, \Big\langle\left[{\cal D}^- - g_{\rmii E}\, {\cal K}^- \right] \phi(y)\, O(z)\Big\rangle &=\expval{\frac{\delta O(z)}{\delta \bar{\phi} (y)}} \label{eq:EOMf}\, ,
\end{align}
for a generic interpolating field $O(z)$, where the differential operators ${\cal D}^\pm$ act
on the $y$ coordinates. Analogous equations hold for $\bar\chi$ and $\bar\phi$. We are interested in the field propagators of
$\chi$ and $\phi$ defined as
\begin{align}
{\cal S}_+(x)=S_\chi(x,0) = \expval{\chi(x)\bar\chi (0)}_f \,, \qquad {\cal S}_-(x) = S_\phi(0,x) = \expval{\phi (0)\bar\phi (x)}_f \,,
\end{align}
where $\expval{\cdot}_f$ refers to the expectation value performed by integrating over the fermionic variables only.
By choosing $O(z)=\bar\chi (z)$, $y=x$ and $z=0$ and $O(z)=\bar\phi(z)$, $y=0$ and $z=y$ in
eqs.~(\ref{eq:EOMc}) and (\ref{eq:EOMf}) respectively, the propagators satisfy the equations of motion
\be
\label{eq:EOM1}
     \Big\langle \left[{\cal D}^+ - g_{\rmii E} \, {\cal K}^+ \right] {\cal S}_+(x)\Big\rangle  = -i\id \delta^{(3)}(x) \, ,\quad 
\Big\langle \left[{\cal D}^+ - g_{\rmii E} \, {\cal K}^- \right]^{\rm T}
{\cal S}^{\rm T}_-(x)\Big\rangle = -i\id \delta^{(3)}(x)\, ,
\ee
where $\id$ is the identity in spinor and colour indices, and the transpose acts on the
same indices. The expectation values in eq. \eqref{eq:EOM1}
are meant to be taken over the gauge field only, however such equations are also valid at fixed gauge
field background.

\subsubsection{Perturbative expansion}
By setting $g_{\rmii E}=0$ in the equations of motion, it straightforward to derive their solutions for the quark
propagators at tree-level given in Appendix~\ref{App:B} which read
\begin{align}
\label{eq:quark_prop_free}
    {\cal S}_\pm^{(0)}(\mathbf{r},x_3) = -i\theta (x_3)\id  \int_{\mathbf{p}} e^{i\mathbf{p}\cdot \mathbf{r}} e^{-x_3 \left( M+\frac{\mathbf{p}^2}{2\pi T} \right)}\, ,
\end{align}
where we have shown separately the dependence on the transverse and longitudinal coordinates.
At the next-to-leading order, by expanding at $O(g_{\rmii E})$, the quark propagators read
\be\label{eq:quark_prop_nlo_1}
    {\cal S}_\pm(\mathbf{r},x_3) = {\cal S}_\pm^{(0)}(\mathbf{r},x_3) + g_{\rmii E} {\cal S}_\pm^{(1)} (\mathbf{r},x_3) + \dots \,,\\[.2cm]
\ee
where
\be
\label{eq:quark_prop_nlo}
    {\cal S}_\pm^{(1)} (\mathbf{r},x_3) =\int_0^{x_3} \diff z_3 \, {\cal K}^{\pm}
    \Big(\frac{z_3}{x_3}\mathbf{r},z_3\Big)\,
    {\cal S}_\pm^{(0)}(\mathbf{r},x_3) \, ,
\ee
and, as in Ref.~\cite{Giusti:2024mwq}, in eq. \eqref{eq:quark_prop_nlo} we have safely taken that,
since quark fields are very heavy, they propagate along the classical trajectory between $({\bf 0},0)$
and $({\bf r},x_3)$.
Notice that, at variance of the computations performed in Ref. \cite{Giusti:2024mwq}, here the quark
propagators at the next-to-leading order are no longer diagonal in Dirac space due to the presence of
the non-diagonal term in ${\cal K}^\pm$.

\section{Mesonic correlators}
\label{sec:mesonic_eft}
The expression for the mesonic interpolating fields in the dimensionally reduced
effective theory can be promptly derived from eq.~\eqref{eq:operators}
by using the conventions in Appendix A of Ref. \cite{Giusti:2024mwq} for the
Clifford algebra and by employing the three-dimensional representation for the quark fields as
in eq. \eqref{eq:NRQCD}. Coherently with the choice made for the action,
we consider the contributions from the $n=0$ fermionic Matsubara
sector only, and we displace the quark fields in the transverse directions as in
Refs.~\cite{Laine:2003bd,Giusti:2024mwq},
so that
\be
\int_0^{1/T} \diff x_0\, {\cal O}^a (x_0,x) \to {\cal O}^a(\mathbf{r}_1,\mathbf{r}_2;x_3)\, ,
\ee
and the point-split fields
read\footnote{We use the same symbol for the fields and correlation functions in QCD and in the effective
theory since the ambiguity is easily resolved from the context.}
\be
{\cal O}^a(\mathbf{r}_1,\mathbf{r}_2;x_3) =
\Big[\bar\chi (\mathbf{r}_1,x_3)
    \,\Sigma_{\cal O}\, T^a \phi (\mathbf{r}_2,x_3) -
    \bar\phi (\mathbf{r}_1,x_3) \,\Sigma_{\cal O}\, T^a \chi (\mathbf{r}_2,x_3) \Big] \,,
\ee
where from passing from QCD to the effective theory
$\Gamma_{{\cal O}}=\left\{\gamma_5,\gamma_2\right\} \rightarrow \Sigma_{\cal O}\ = \left\{ \sigma_3,\sigma_1 \right\}$
for pseudoscalar and vector fields respectively. These fields are not gauge invariant for
$\mathbf{r}_1\neq \mathbf{r}_2$. However, as we will see in the following, the results for the
screening masses will be gauge independent, as one can easily prove that gauge dependent contributions in
their correlation functions vanish in the large $x_3$ limit. 
In the effective field theory, the two-point screening correlation functions introduced in eq. \eqref{eq:2pt}
map to
\begin{align}
    {\cal C}_{\cal O} (x_3) = \int_{\mathbf{R}} {\cal C}_{\cal O} (\mathbf{r}_1,\mathbf{r}_2; x_3)\Big\rvert_{\mathbf{r}_1=\mathbf{r}_2=\mathbf{R}} = T \int_{\mathbf{R}} \expval{{\cal O}^a(\mathbf{r}_1,\mathbf{r}_2;x_3) {\cal O}^a(0)}\Big\rvert_{\mathbf{r}_1=\mathbf{r}_2=\mathbf{R}} \,.
\end{align}
When performing the integration over the fermionic variables and taking the Wick contractions, the quark propagators are diagonal
in flavour space. As a consequence, by making explicit the flavour indices, the flavour structure simplifies
to $\Tr[T^a T^b]=\delta^{ab}/2$. By considering the latter, and the factor $2$ which takes
into account the contribution from the $n=-1$ Matsubara sector not explicitly indicated in the
previous formulas, the general expression for the correlation functions is
\begin{align}
\label{eq:wick}
    {\cal C}_{\cal O}(\mathbf{r}_1,\mathbf{r}_2;x_3) = T \expval{\Tr \Big[ \Sigma_{\cal O} \,  {\cal S}_+(\mathbf{r}_1,x_3)
    \, \Sigma_{\cal O} \, {\cal S}_-(\mathbf{r}_2,x_3)\Big]} = T\, \Big\langle W_{\cal O}(\mathbf{r}_1,\mathbf{r}_2;x_3)\Big\rangle 
\end{align}
where the trace is over colour and spinor indices.

\subsection{Free-theory limit}
The expression of the Wick contraction in the free theory is derived by inserting the
propagators in eq. \eqref{eq:quark_prop_free} into eq. \eqref{eq:wick}. By contracting colour
and spinor indices, the Wick contraction,
independently of the matrix $\Sigma_{\cal O}$, can be written as
\begin{align}
\label{eq:W0}
    \begin{aligned}
        W_{{\cal O}}^{(0)}(\mathbf{r}_1,\mathbf{r}_2;x_3) & = 
        \Tr\Big[{\cal S}^{(0)}_+(\mathbf{r}_1,x_3) {\cal S}^{(0)}_-(\mathbf{r}_2,x_3)\Big]  \\
        &= - 2 N_c  \, \theta (x_3)\int_{\mathbf{p}_1,\mathbf{p}_2} e^{i\left( \mathbf{p}_1\cdot \mathbf{r}_1 + \mathbf{p}_2\cdot\mathbf{r}_2\right)} e^{-x_3 \left( 2M+\frac{\mathbf{p}_1^2}{2\pi T}+\frac{\mathbf{p}_2^2}{2\pi T} \right)} \, ,
    \end{aligned}
\end{align}
and it satisfies the (2+1)-dimensional Schr\"odinger equation
\begin{align}
    \left[ 2M +\partial_3 -\sum_{i=1,2} \frac{\nabla_{\mathbf{r}_i}^2}{2\pi T} \right] W_{{\cal O}}^{(0)}(\mathbf{r}_1,\mathbf{r}_2;x_3) \overset{x_3>0}{=}0 \, .
\end{align}
 It follows that, by assuming heavy quarks with small transverse momentum, i.e. longitudinal propagation, the exponential fall-off
 of $W_{\cal O}^{(0)}(\mathbf{r}_1,\mathbf{r}_2;x_3)$ is dominated by $2M=2\pi T+ O(g^2)$ which is therefore the leading contribution to both the pseudoscalar and the vector screening masses.

\subsection{Next-to-leading order}
The next-to-leading order spin-independent correction to the flavour non-singlet mesonic screening masses has been
computed, in the framework of the dimensionally-reduced effective theory, in Ref. \cite{Laine:2003bd}, see
eq.~\eqref{eq:m_nlo}. Here such calculation is extended in order to obtain the leading
spin-dependent correction
to the screening masses. Starting from the equations of motion for the quark propagators in eq.~\eqref{eq:EOM1},
the equation of motion for $\langle W_{\cal O}\rangle$ in the interacting case reads
\begin{align}
\label{eq:EOM-next-to-leading}
\begin{aligned}
 & \Biggl[ 2M + \partial_3 -\sum_{i=1,2} \frac{\nabla_{\mathbf{r}_i}^2}{2\pi T} \Biggr]
    \Big\langle W_{\cal O}(\mathbf{r}_1,\mathbf{r}_2;x_3)\Big\rangle \overset{x_3>0}{=}\\
 &  \hspace{1.0cm}  \overset{x_3>0}{=} g_{\rmii E} \biggl \langle
    \Tr \Big\{\Big[{\cal A}^+(\mathbf{r}_1,x_3)+{\cal A}^-(\mathbf{r}_2,x_3)\Big]
 \Sigma_{\cal O} {\cal S}_+ (\mathbf{r}_1,x_3) \Sigma_{\cal O} {\cal S}_-\, (\mathbf{r}_2,x_3) \Big\}\\ 
&   \hspace{1.25cm} - \frac{1}{2\pi T} \Tr \Big\{\Big[s_{\cal O} B_3(\mathbf{r}_1,x_3) + B_3(\mathbf{r}_2,x_3)\Big]
 \sigma_3\, \Sigma_{\cal O}   {\cal S}_+(\mathbf{r}_1,x_3)\, \Sigma_{\cal O}\, {\cal S}_-(\mathbf{r}_2,x_3)\Big\}
       \biggr \rangle \,  .
\end{aligned}
\end{align}
where ${\cal A}^\pm$ is defined in eq. \eqref{eq:Apm}, and we have introduced the notation
$s_{\cal O}=(+1,-1)$ for pseudoscalar and vector respectively, see Appendix~\ref{App:C}.
By using the quark propagators at $O(g_{\rmii E})$ in eq. \eqref{eq:quark_prop_nlo_1}, and by performing the
gluon contractions (see Appendix \ref{App:C} for further details) we obtain
\be
     \Biggl[ 2M + \partial_3 - \sum_{i=1,2} \frac{\nabla_{\mathbf{r}_i}^2}{2\pi T}  \Biggr]
     \Big\langle W_{\cal O}(\mathbf{r}_1,\mathbf{r}_2;x_3)\Big\rangle  \overset{x_3>0}{=}\! -\, 
     {\cal U}(\mathbf{r}_1,\mathbf{r}_2;x_3)\, W_{\cal O}^{(0)}(\mathbf{r}_1,\mathbf{r}_2;x_3)  
     +  O(g_{\rm E}^3)\, , \label{eq:EOM-next-to-leading-1}
\ee
where the potential is 
\be\label{eq:USI}
    {\cal U}(\mathbf{r}_1,\mathbf{r}_2;x_3) =  {\cal U}_{{\rm SI}_1}(\mathbf{r}_1,\mathbf{r}_2;x_3) +
    {\cal U}_{{\rm SI}_2}(\mathbf{r}_1,\mathbf{r}_2;x_3) + {\cal U}_{\cal O}(\mathbf{r}_1,\mathbf{r}_2;x_3)\, ,
\ee
with
\bea
{\cal U}_{{\rm SI}_1}(\mathbf{r}_1,\mathbf{r}_2;x_3) & = &  -\frac{g_{\rmii E}^2}{N_c}
    \Big\langle\Tr\Big\{\Big[ {\cal A}^+(\mathbf{r}_1,x_3) + {\cal A}^-(\mathbf{r}_2,x_3)\Big] \times\label{eq:USI1}\\[0.125cm]
    & & \hspace{2.25cm} \int_0^{x_3} \diff z_3
    \Big[ {\cal A}^+\Big(\frac{z_3}{x_3}\mathbf{r}_1,z_3\Big) +
    {\cal A}^-\Big(\frac{z_3}{x_3} \mathbf{r}_2,z_3\Big)\Big] \Big\} \Big\rangle\, , \nonumber\\[0.125cm]
{\cal U}_{{\rm SI}_2}(\mathbf{r}_1,\mathbf{r}_2;x_3) &=& -\frac{g_{\rmii E}^2}{(2\pi T)^2 N_c}
\times \label{eq:USI2}\\[0.125cm]
& &\hspace{-1.5cm}     \int_0^{x_3} \diff z_3
 \Big\langle\Tr\Big\{B_3(\mathbf{r}_1,x_3)B_3\Big(\frac{z_3}{x_3}\mathbf{r}_1,z_3\Big) +
                              B_3(\mathbf{r}_2,x_3)B_3\Big(\frac{z_3}{x_3}\mathbf{r}_2,z_3\Big)
\Big\}\Big\rangle , \nonumber
\eea
and the chromo-magnetic field $B_3$ is defined at leading order in eq.~\eqref{eq:magnetic_field}.
The ${\cal U}_{{\rm SI}_1}$ potential is the one that was obtained in Ref.~\cite{Laine:2003bd}.
The second spin-independent
contribution ${\cal U}_{{\rm SI}_2}$ is temperature suppressed. It derives from the exchange of
longitudinal ultrasoft gluons along the same quark line.
The leading contribution to the spin-dependent potential reads
\bea
{\cal U}_{\cal O}(\mathbf{r}_1,\mathbf{r}_2;x_3) & = & -
\frac{g_{\rmii E}^2  s_{\cal O}}{(2\pi T)^2 N_c}\times \label{eq:US}\\[0.125cm]
& &\hspace{-1.5cm} \int_0^{x_3} \diff z_3
\Big\langle\Tr\Big\{B_3(\mathbf{r}_1,x_3)B_3\Big(\frac{z_3}{x_3}\mathbf{r}_2,z_3\Big) + 
B_3(\mathbf{r}_2,x_3)B_3\Big(\frac{z_3}{x_3}\mathbf{r}_1,z_3\Big) \Big\} \Big\rangle \,.\nonumber
\eea
It is temperature suppressed, and it is due to the exchange of ultrasoft gluons
between two quark lines. Since to extract the screening masses we are interested in the large
$x_3$ behaviour of the Wick contractions in eq.~\eqref{eq:wick}, 
we take the limit $x_3\to \infty$ in
eq.~(\ref{eq:EOM-next-to-leading-1}) which reads
\be
\label{eq:Schrodinger1}
    \Biggl[2M + \partial_3 -\sum_{i=1,2} \frac{\nabla_{\mathbf{r}_i}^2}{2\pi T}
    + U(\mathbf{r}_1-\mathbf{r}_2)\Biggr]
    \Big\langle W_{\cal O}(\mathbf{r}_1,\mathbf{r}_2;x_3) \Big\rangle = 0
    + O(g_{\rm E}^3) \,,
\ee
where 
\be
U(\mathbf{r}_1-\mathbf{r}_2) =  \lim_{x_3\to \infty}
\Big[ {\cal U}_{{\rm SI}_1}(\mathbf{r}_1,\mathbf{r}_2;x_3) +
{\cal U}_{{\rm SI}_2}(\mathbf{r}_1,\mathbf{r}_2;x_3) + {\cal U}_{\cal O}(\mathbf{r}_1,\mathbf{r}_2;x_3) \Big] \, ,\\
\ee
and, at the order we work, we could replace $W^{(0)}_{\cal O} \to \expval{W_{\cal O}}$.

\subsubsection{Spin-independent contribution\label{eq:SIsec}}
 The spin-independent contribution ${\cal U}_{{\rm SI}_2}$ in eq.~\eqref{eq:USI2} is temperature-suppressed with respect to the one
 in eq.~\eqref{eq:USI1}, and it can be neglected when keeping terms up to $O(g_{\rmii E}^2/T)$ only. By taking into account the discussion
 in Section~\ref{sec:results}, this potential would produce subleading contributions in the
 wave-functions of the screening mass states, and therefore in the hyperfine splitting. Similarly, since we are interested in the
 leading contribution to eq.~\eqref{eq:DeltaM} only, we can safely take the expression of the low energy constant $M$ in eq.~(\ref{eq:M}).
 By taking the large separation limit of the potential in eq. \eqref{eq:USI1}, see Appendix \ref{App:D} for the intermediate steps,
 we obtain for the next-to-leading spin-independent contribution
\begin{align}
\label{eq:static_SI}
    U_{{\rm SI}_1}(\mathbf{r}) = \lim_{x_3\to \infty} {\cal U}_{{\rm SI}_1}(\mathbf{r}_1,\mathbf{r}_2;x_3) = \frac{g_{\rmii E}^2 C_F}{2\pi} \left[ \ln \left( \frac{m_{\rmii E}r}{2}\right) +\gamma_{\rmii E} -K_0\left( m_{\rmii E} r \right) \right]
\end{align}
where $\mathbf{r}=\mathbf{r}_1-\mathbf{r}_2$ and $r=|\mathbf{r}|$, $\gamma_{\rmii E}$ is the
Euler-Mascheroni constant and $K_0$ is a modified Bessel function. Note that the leading logarithmic
behaviour on the r.h.s. of eq.~(\ref{eq:static_SI}) is a confining Coulomb interaction in $2+1$ dimensions.
By combining the power counting reported in Ref. \cite{Giusti:2024mwq} and standard dimensional analysis
arguments, it is possible to see that a string term, i.e. a non-perturbative confining term, arises in the potential
at $O(g^3)$. This limits, de facto, the applicability of the perturbative approach for computing
the spin-independent potential to the $O(g^2)$.

\subsubsection{Spin-dependent contribution}
The explicit expression for the spin-dependent static potential $U_{\cal O}$ in eq.~\eqref{eq:US}
is carried out analogously, see
Appendix~\ref{App:D} for the intermediate steps and the detailed calculations. By taking the large separation limit in the
longitudinal direction, it yields to
\begin{align}
\label{eq:static_SD}
    U_{\cal O}(\mathbf{r}) = \lim_{x_3\to \infty} {\cal U}_{\cal O}(\mathbf{r}_1,\mathbf{r}_2;x_3) =
    - g_{\rmii E}^2 \frac{s_{\cal O} C_F}{(2\pi T)^2} \delta^{(2)}(\mathbf{r}) \, .
\end{align}
Notice that, while it provides an interaction which is of the same order in the effective coupling constant
$g_{\rmii E}$ with respect to the spin-independent potential in eq. \eqref{eq:static_SI}, it is however
temperature-suppressed.

\section{Hyperfine splitting}
\label{sec:results}
As we have seen in the previous section, the equation of motion for the two-point mesonic screening correlator
in the three-dimensional effective theory implies the Schr\"odinger equation in eq. \eqref{eq:Schrodinger1},
and the screening mass corresponds to its ground-state energy. Since the spin-dependent potential
in eq. \eqref{eq:static_SD} provides a temperature-suppressed contribution with
respect to the leading spin-independent potential in eq. \eqref{eq:static_SI}, $U_{\cal O}({\bf r})$
can be treated as 
a perturbative correction. We then start by solving the
unperturbed Schr\"odinger equation, which, in the large $x_3$ limit, reads
\begin{align}
\label{eq:Schrodinger2}
    \left[ - \frac{\nabla_{\mathbf{r}_1}^2+\nabla_{\mathbf{r}_2}^2}{2\pi T} + 2M + U_{\rm{SI}_1}(\mathbf{r}_1-\mathbf{r}_2) \right] \Psi_0(\mathbf{r}_1,\mathbf{r}_2) = E_0 \Psi_0(\mathbf{r}_1,\mathbf{r}_2) \, ,
\end{align}
where, with $\Psi_0$ and $E_0$, we refer to the ground-state energy and wave-function at leading order
in perturbation theory\footnote{In this section we omit any label referring to the eigenstates of the
Hamiltonian since we are only interested in the ground state of the system.}.
By going to the center-of-mass frame, we define
\be    
    \mathbf{R} =  \, \frac{\mathbf{r}_1+\mathbf{r}_2}{2} \, ,\qquad \mathbf{r} =  \,  \mathbf{r}_1-\mathbf{r}_2 \,,\\ 
\ee
and the Laplace operator can accordingly be written as
\begin{align}
    \nabla_{\mathbf{r}_1}^2+\nabla_{\mathbf{r}_2}^2 = \frac{1}{2}\nabla_{\mathbf{R}}^2 + 2\nabla_{\mathbf{r}}^2 \, ,
\end{align}
where the first term on the r.h.s. describes the motion of the center-of-mass, while the second one
is the kinetic term related to the relative motion. By restricting to the relative motion only,
and by making explicit the functional form of the potential, see eq.~\eqref{eq:static_SI}, the spin-independent
Schr\"odinger equation reads
\begin{align}
\label{eq:Schrodinger}
    \left\lbrace -\frac{\nabla_{\mathbf{r}}^2}{\pi T} + 2M + g_{\rmii E}^2 \frac{C_F}{2\pi} \left[ \ln \left(\frac{m_{\rmii E}r}{2}\right)+\gamma_{\rmii E} -K_0(m_{\rmii E}r)   \right] - E_{0} \right\rbrace \psi_0(\mathbf{r}) = 0 \, ,
\end{align}
This is exactly the Schr\"odinger equation obtained in Ref. \cite{Laine:2003bd} which leads the
next-to-leading order value of the mesonic screening mass in eq. \eqref{eq:m_nlo}.
Once $E_0$ and the normalized $\psi_0$ have been determined for the unperturbed Schr\"odinger equation,
the mesonic screening masses at the first order in the spin-dependent perturbative expansion
reads
\be\label{eq:Ecorrection}
\displaystyle
E_{\cal O} = E_0 +  \int_{\mathbf{r}} U_{\cal O}({\bf r})\big|\psi_0({\bf r})\big|^2
         =  E_0 - g_{\rmii E}^2 \frac{s_{\cal O} C_F}{(2\pi T)^2} \big|\psi_0({\bf 0})\big|^2\,.
\ee
It is important to notice that, even though the spin-dependent potential is localized in $\mathbf{r}=0$,
the hyperfine
correction to the screening masses receives contributions from large distances through
the normalization of the wave-function, see discussion below.

\subsection{Leading contribution}
The solution of the spin-independent eigenvalue problem for the ground state is found by going to polar coordinates,
and by considering the case of vanishing angular momentum. By defining
$\hat\psi_0(\hat{r}) = \psi_0({\bf r}) \sqrt{2\pi}/m_{\rmii E}$, where we introduced the dimensionless variable
$\hat{r}=m_{\rmii E}r$, Eq. \eqref{eq:Schrodinger2} can then be rewritten as
\begin{align}
    \label{eq:Schrodinger-hat}
    \left\lbrace -\left( \frac{\diff^2}{\diff \hat{r}^2} + \frac{1}{\hat{r}} \frac{\diff}{\diff \hat{r}} \right) + \rho \left[ \ln \left(\frac{\hat r}{2}\right)+\gamma_{\rmii E} -K_0(\hat r) - \hat{E}_0  \right]
    \right\rbrace \hat\psi_0(\hat{r}) = 0 \, .
\end{align}
Accordingly to Ref. \cite{Laine:2003bd},
we introduced the parameterization of the effective coupling
\begin{align}
    \rho = \frac{g_{\rmii E}^2 C_F T}{2m_{\rmii E}^2} = \frac{4}{9} \, , 
\end{align}
where in the last step we used $N_f=3$, $N_c=3$ and the expressions at leading order for
$m_{\rmii E}$ and $g_{\rmii E}^2$ reported in eq. \eqref{eq:ge}. Furthermore we introduced the
dimensionless energy eigenvalue \cite{Laine:2003bd}
\begin{align}
   \hat{E}_0 =  \frac{2\pi}{g_{\rmii E}^2 C_F} \left( E_0 - 2M \right) \,.
\end{align}
By doing the same change of variable in the matrix element of the spin-dependent potential and in the normalization of the wave-function, the spin-dependent eigenvalue in eq. \eqref{eq:Ecorrection} can be written as
\begin{align}
\label{eq:E_sigma}
    E_{\cal O} = E_0 - \frac{g_{\rmii E}^2 m_{\rmii E}^2}{2\pi} \frac{s_{\cal O} C_F}{(2\pi T)^2} |\hat\psi_0(0)|^2 \, ,
\end{align}
where
\be
\int_0^{\infty} d\hat{r} \, \hat{r} \, |\hat\psi_0(\hat r)|^2 = 1\, .
\ee
By recalling that $s_{\cal O}=(+1,-1)$ for the pseudoscalar and the vector correlators
respectively, and by inserting the expression at leading order for
$m_{\rmii E}^2$ and $g_{\rmii E}^2$, see eq. \eqref{eq:ge}, it follows that
the hyperfine-splitting in the pseudoscalar and vector energy levels due to spin-dependent
interactions at leading order in perturbation theory, cf. eq. \eqref{eq:DeltaM},
is
\begin{align}
\label{eq:HF}
    \frac{\Delta m_{\rmii{VP}}^{\rmii{lo}}}{2\pi T} = g^4 \left(\frac{N_c}{3}+\frac{N_f}{6} \right)
    \frac{C_F}{8\pi^4} |\hat \psi_0(0)|^2 \,,
\end{align}
where we substituted $E_{\rmii P} = m_{\rmii P}$ and $E_{\rmii V}=m_{\rmii V}$ in eq. \eqref{eq:E_sigma}.
The numerical solution
for $\hat \psi_0$ of the unperturbed Schr\"odinger equation (\ref{eq:Schrodinger-hat}) gives
\begin{align}
\label{eq:HF-splitting}
     \frac{\Delta m_{\rmii{VP}}^{\rmii{lo}}}{2\pi T} =  0.002376 \cdot g^4\, .
\end{align}
This result justifies a posteriori the choice of neglecting $O(g_{\rmii E}^2/T^2)$ terms in the spin-independent 
potential in Section~\ref{eq:SIsec}, since it is clear that the inclusion of the potential obtained
from eq. \eqref{eq:USI2} into eq.~(\ref{eq:Schrodinger2}) would have produced higher order effects in the
spin-splitting between the pseudoscalar and the vector screening masses. A similar discussion holds for the
higher order contributions deriving from the perturbative matching between the three-dimensional non-relativistic
theory in eq. \eqref{eq:NRQCD} and QCD.
From the discussion in Section~\ref{eq:SIsec}, a string term of non-perturbative origin
starts at $O(g^3)$ in the static potential, and, as a consequence, non-perturbative effects are expected to
start contributing to the hyperfine splitting at $O(g^5)$. Therefore, to go beyond the result in
eq.~(\ref{eq:HF}), a non-perturbative approach is required.

\subsection{Analysis of the wave-function\label{sec:wave_function_analysis}}
\begin{figure}[t]
    \centering
    \includegraphics[width=0.33\linewidth]{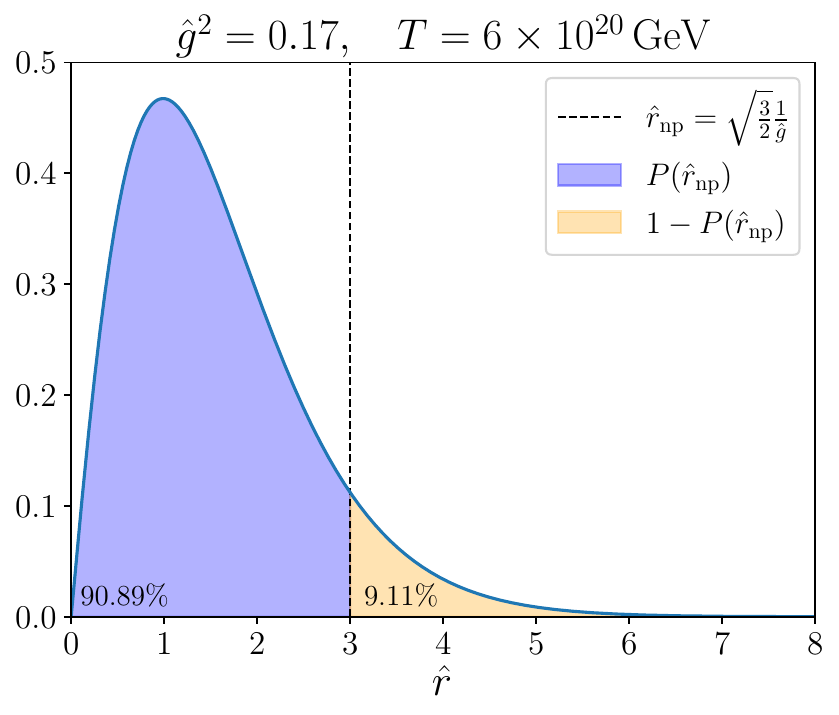}%
    \includegraphics[width=0.33\linewidth]{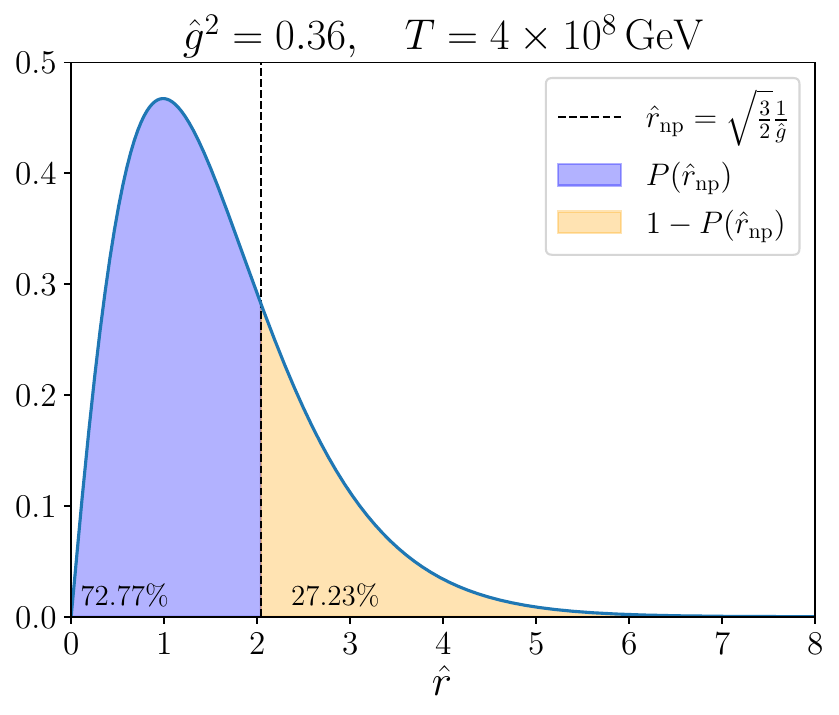}%
    \includegraphics[width=0.33\linewidth]{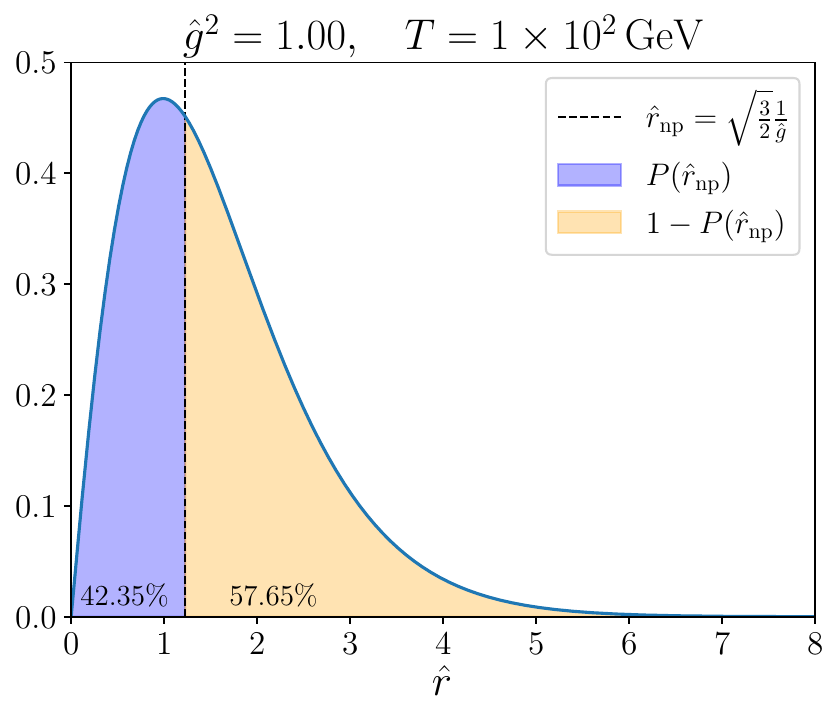}
    \caption{Probability density for the quark-antiquark pair as a function of
             the distance $\hat r$. The probability for the quark and the antiquark
             to be at distances $\hat{r} \leq \hat{r}_{\rm{np}}$, see
             eq.~(\ref{eq:probability}), is indicated by the blue area for
             different values of the coupling constant (temperature), with the latter
             increasing from left to right. As the temperature decreases, the blue area
             becomes larger and larger.}
    \label{fig:probability}
\end{figure}

The perturbative result for the hyperfine-splitting in eq.~(\ref{eq:HF}) probes the
normalized wave function of the quark-antiquark bound system at the origin. In particular its
normalization takes contributions from all distances, and therefore from both the scales
present in EQCD, i.e. $r \sim m^{-1}_\rmii{E}$ and $(g^2 T)^{-1}$. To scrutinize this issue, it is interesting
to define the probability
\begin{align}
\label{eq:probability}
    P(\hat{r}_{\rm np}) = \int_0^{\hat{r}_{\rm np}} d\hat{r} \, \hat{r}
    \big|\hat \psi_0(\hat{r})\big|^2 \, ,
\end{align}
for the quark-antiquark pair to be at a distance $\hat r\leq \hat r_{\rm np}$, where  
\begin{align}
    \hat{r}_{\rm np} = m_{\rmii E} r_{\rm np} = \frac{m_{\rmii E}}{g^2_{\rmii E}} = \left(\frac{N_c}{3}+\frac{N_f}{6} \right)^{1/2} \frac{1}{g} = \sqrt{\frac{3}{2}} \frac{1}{g} \, ,
\end{align}
and on the r.h.s.~we substituted $N_f=3$ and $N_c=3$ which is the relevant case to this study.
In Figure~\ref{fig:probability} we show the probability density as a function of $\hat{r}$ together
with the ratio $\hat{r}_{\rm np}$ of the two relevant scales (dashed line).
As expected, at asymptotically high temperatures (left panel) the probability for the quark and the antiquark
to be at distances $\hat{r} \leq \hat{r}_{\rm{np}}$ is close to $1$ (blue band), and the
perturbative determination in eq. \eqref{eq:HF} becomes more and more reliable as
the temperature increases. In the central and right panels of Figure~\ref{fig:probability}
the cases for $T \sim 4\times 10^8$~GeV ($P(\hat{r}_{\rm np})\sim 70\%$) and 
for $T \sim \times 10^2$~GeV ($P(\hat{r}_{\rm np})\sim 40\%$) are shown, respectively.
It is rather clear
that, at temperatures of the order of the electro-weak scale, the probability for the quark and the antiquark
to be at distances larger than  $\hat{r}_{\rm{np}}$ approaches $50\%$. As a consequence,
long-distance (ultra-soft) contributions are relevant, and non-perturbative effects must be taken into account.
As discussed previously, they start contributing already at $O(g^5)$
for the hyperfine splitting.\\
\indent This analysis, albeit rather crude, is consistent with the fact that the hierarchy of
scales in eq.~(\ref{eq:scales}) sets in only at much higher temperatures with respect to the  electro-weak
scale, where $g\sim 1 $. The result in eq.~(\ref{eq:HF}), however, remains extremely useful
to discriminate, among the possible interpolating fit functions of the non-perturbative data, those which have the
correct asymptotic behaviour at asymptotically large temperatures.

\section{Comparison with non-perturbative results}
\label{sec:comparison}
The flavour non-singlet pseudoscalar and vector screening masses have been computed
recently in thermal QCD with $N_f=3$ massless quarks in the range of temperatures from
$T\sim 1$~GeV up to $\sim 160$~GeV with high precision~\cite{DallaBrida:2021ddx}.
\begin{figure}
    \centering
    \includegraphics[width=0.48\linewidth]{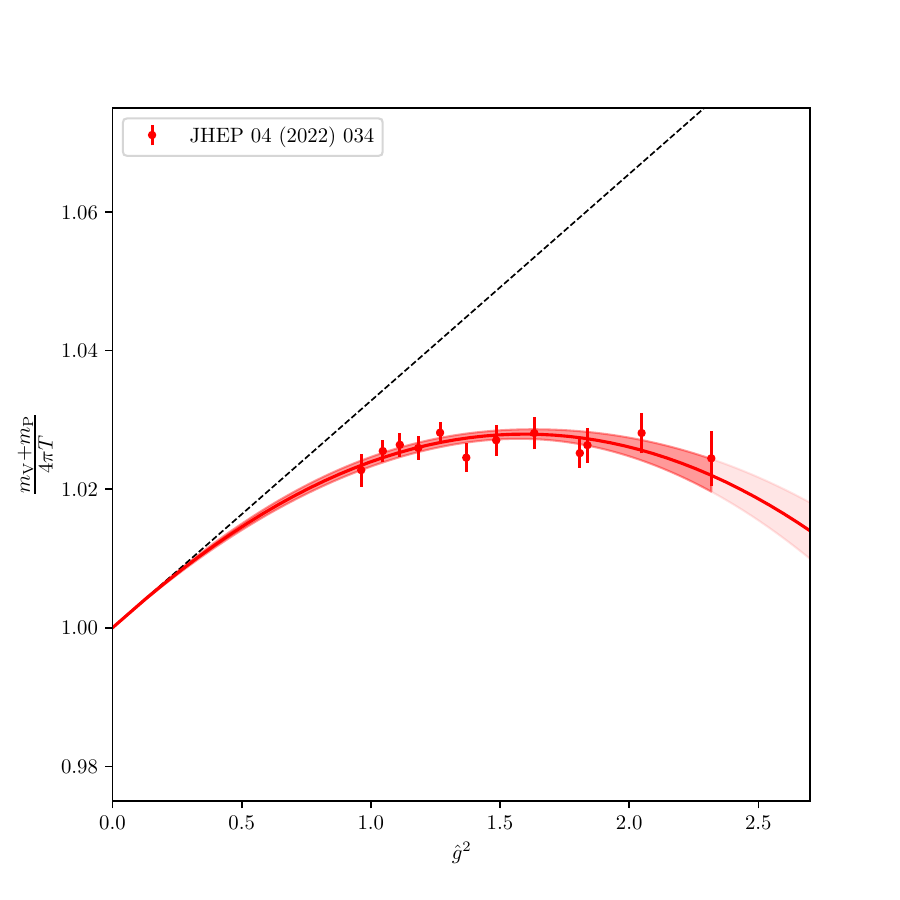}
    \includegraphics[width=0.48\linewidth]{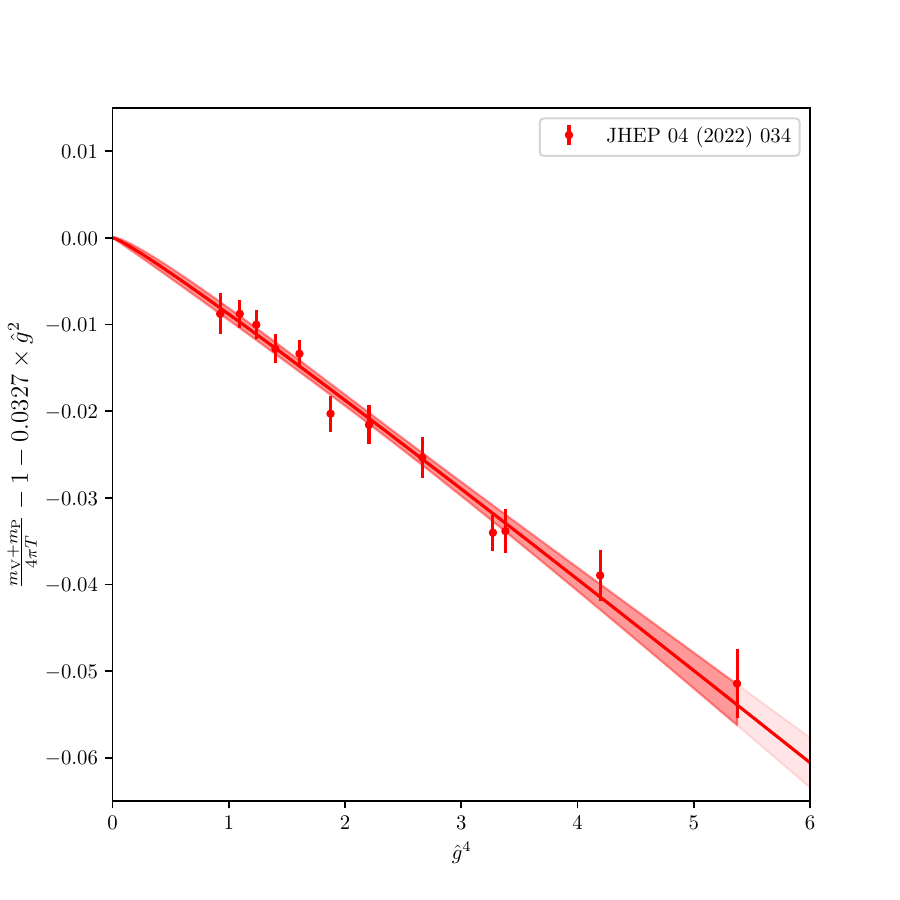}
    \caption{Left: sum of the pseudoscalar and the vector screening masses, normalized to $4\pi T$, as a function of
      $\hat{g}^2$. Lattice data are obtained from Ref. \cite{DallaBrida:2021ddx}, the red band represents the best
      parameterization in eq.~\eqref{eq:parameterization_sum} with its error, while the black dashed line is the
      analytically known spin-independent contribution in eq. \eqref{eq:m_nlo}. Right: same as in the left panel
      but with the lattice data, subtracted by the analytically known contribution, plotted versus $\hat{g}^4$.}
    \label{fig:sum}
\end{figure}
By remembering that $m_{\rmii P}=E_{\rmii P}$ and $m_{\rmii V}=E_{\rmii V}$, where the expressions for
$E_{\rmii P}$ and $E_{\rmii V}$ are given in
eq. \eqref{eq:E_sigma}, it follows that the magnitude of the leading spin-dependent
correction is the same for both masses, but with opposite sign. Therefore, up to $O(g^4)$, the sum of the
pseudoscalar and the vector
screening masses probes the spin-independent potential only. In the two panels of Figure \ref{fig:sum} we report
the non-perturbative data for $(m_{\rmii V}+m_{\rmii P})/4\pi T$ from Ref.~\cite{DallaBrida:2021ddx} as a function of $\hat{g}$,
the renormalized coupling in the $\overline{\rm MS}$ scheme at two-loop order evaluated at the renormalization
scale $\mu=2\pi T$, see eq. (7.1) of Ref.~\cite{DallaBrida:2021ddx} for its precise definition.
We stress once again that, for our purposes,
$\hat{g}$ is just a function of the temperature $T$, suggested by the effective theory analysis, that we use
to analyze the non-perturbative data. The crucial point is its leading logarithmic dependence on $T$. In
the left panel of Figure~\ref{fig:sum} the bending down of the normalized sum of the the two masses
as the temperature decreases signals the presence of higher order contributions with respect to the
leading interacting term in eq. \eqref{eq:m_nlo}. Data exhibit a very mild temperature dependence
which is originated by the competition between the leading contribution in eq. \eqref{eq:m_nlo} and
higher order corrections. This results in a positive correction of about $2\%$ with respect to the free
field theory result in the entire range  of temperature explored. In the right panel of Figure~\ref{fig:sum}
we show the normalized sum of the two masses after subtracting the known leading contribution in eq.~\eqref{eq:m_nlo},
plotted as a function of $\hat{g}^4$. At the highest temperature, i.e. $\hat{g}^4\sim 1$, higher order terms
provide a contribution which is still about $30\%$ of the leading interacting term, while at $T\sim 1$ GeV,
i.e. $\hat{g}^4\sim 5$, this contribution increases up to $\sim 70\%$. Given these considerations, we parameterize
the non-perturbative data with the polynomial 
\begin{align}
\label{eq:parameterization_sum}
    \frac{m_{\rmii V}+m_{\rmii P}}{4\pi T} = p_0 + p_2 \hat{g}^2 + p_3\hat{g}^3 + p_4\hat{g}^4 \, .
\end{align}
A fit of the data reveals that the leading and the quadratic coefficients $p_0$ and $p_2$ turns out to be compatible
with the free field theory and the leading perturbative interacting contributions respectively, see eq. \eqref{eq:m_nlo}.
We thus enforce $p_0=1$ and $p_2=0.032740$, and fit the data again. As a result, we obtain, for the remaining fit parameters,
$p_3=0.0036(29)$ and $p_4=-0.0124(23)$, with ${\rm{cov}}(p_3,p_4)/\left[\sigma(p_3)\sigma(p_4) \right]=-0.99$ and
$\chi^2/{\rm{d.o.f.}}=0.60$. Given the high quality of this fit, we take this as final parameterization of the lattice data,
and we display it as a red band in both panels of Figure~\ref{fig:sum}. For completeness, let us notice that, by enforcing
$p_3=0$, we obtain $p_4=-0.00959(27)$ and, again, an excellent $\chi^2/{\rm{d.o.f.}}=0.69$. These findings show that the 
data in Ref.~\cite{DallaBrida:2021ddx} are compatible with the expected behaviour at asymptotically large temperatures, but
higher order terms in $\hat g$ are required to explain their temperature dependence.
Since non-perturbative contributions are expected
already at $O(g^3)$, they have to be taken into account for explaining the results in Figure~\ref{fig:sum}
up to temperatures of the order of the electro-weak scale. 

\begin{figure}
    \centering
    \includegraphics[width=0.5\linewidth]{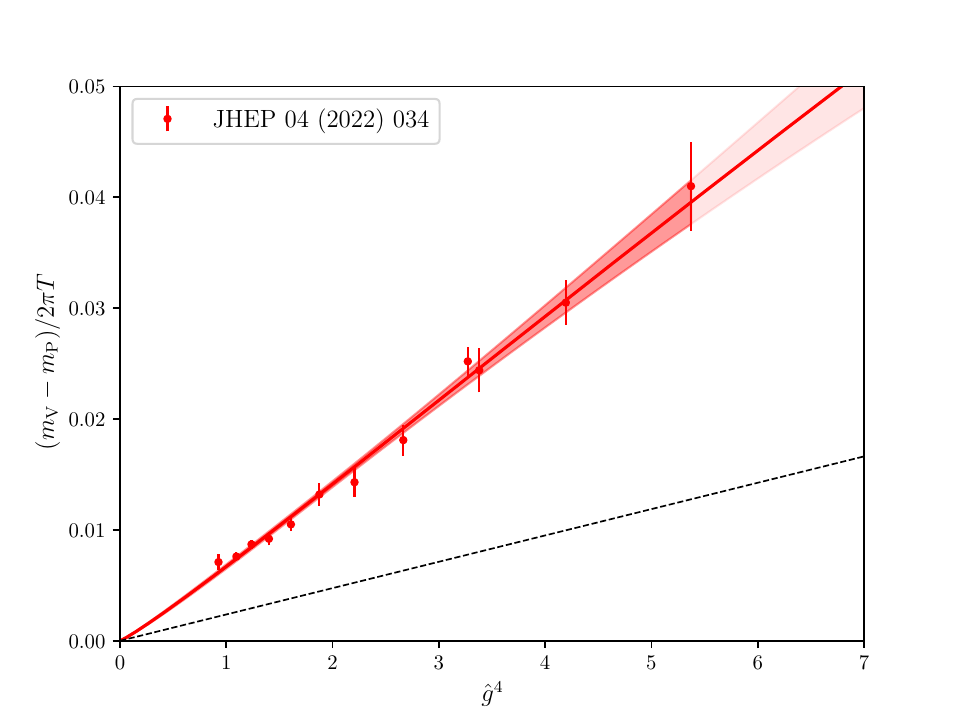}
    \caption{Difference of the vector and pseudoscalar screening masses normalized to $2\pi T$ versus $\hat g^4$. The
      perturbative result in eq. \eqref{eq:HF-splitting} is the dashed black line, while the red curve represents
      the fit in eq. \eqref{eq:DeltaM_fit}.}
    \label{fig:Tdep}
\end{figure}

The spin-dependent contribution to the mesonic screening masses has been determined in Ref.~\cite{DallaBrida:2021ddx}
from the mass difference of $m_{\rmii V}$ and $m_{\rmii P}$. The results are reported in Figure~\ref{fig:Tdep} as a function of
$\hat{g}^4$. Within statistical errors, data show an effective quartic dependence on $\hat{g}$ in the entire range
of temperature explored. However, the effective slope, $0.00704(14)$, turns out to be approximatively $3$ times the
coefficient on the r.h.s. of eq.~(\ref{eq:HF-splitting}). Thanks to the perturbative result in eq.~\eqref{eq:HF-splitting},
however, we
can constrain the asymptotic behaviour of the mass difference between the vector and the pseudoscalar screening masses
at asymptotically large temperatures. We then parameterize the non-perturbative data in
Ref. \cite{DallaBrida:2021ddx} as
\begin{align}
\label{eq:DeltaM_fit}
    \frac{m_{\rmii V}-m_{\rmii P}}{2\pi T} = 0.002376\, \hat g^4 + s_5\, \hat{g}^5 + s_6\, \hat{g}^6\,,
\end{align}
where a fit to the data gives $s_5=0.0063(9)$, $s_6=-0.0020(7)$,
${\rm{cov}}(s_5,s_6)/\left[ \sigma(s_5) \sigma(s_6) \right] = -0.99$, and an excellent $\chi^2/{\rm{d.o.f.}}=0.75$.
The eq.~(\ref{eq:DeltaM_fit}) represents our best parameterization of the non-perturbative data.
Being $s_5$ about $3$ times in magnitude with respect to $s_4$, it follows that the leading $O(\hat{g}^4)$ term
accounts only for a small fraction of the total contribution  in the entire range of temperatures studied
in Ref.~\cite{DallaBrida:2021ddx}. The bulk of the hyperfine splitting originates from higher order
terms which can be present both in the spin-dependent potential and in the unperturbed,
spin-independent wave-function. As a result contributions of non-perturbative origin has to be
taken into account to explain the observed magnitude of the hyperfine splitting up to the
electro-weak scale. This does not come as a surprise given the analysis
in section~\ref{sec:wave_function_analysis}.

\section{Conclusions}
\label{sec:conc}
Recent progress in lattice QCD has made it possible to study the strongly
interacting quark gluon plasma non-perturbatively up to temperatures of the
order of the electro-weak scale. First results indicate that non-perturbative
contributions needs to be considered up to temperatures orders of magnitude larger than
the GeV scale to explain the dynamics of the theory. Of particular interest is
the hyperfine splitting in the QCD flavour non-singlet mesonic screening masses
because it can be computed very precisely on the lattice and it is expected to be
particularly sensitive to non-perturbative dynamics.\\
\indent Motivated by
the results in Ref.~\cite{DallaBrida:2021ddx}, here we have computed analytically
the leading contribution of $O(g^4)$ to the hyperfine splitting at asymptotically large
temperatures. Non-perturbative data indeed show and effective $O(g^4)$
behaviour~\cite{DallaBrida:2021ddx}, within statistical errors,
in the temperature range from $T\sim 1$~GeV up to $~160$~GeV. The effective prefactor, however,
is approximatively $3$ times larger than the one of the leading $O(g^4)$ contribution at asymptotically
high temperatures.
By enforcing the $O(g^4)$ term to its analytic value in eq.~(\ref{eq:HF-splitting}), non-perturbative data
for the hyperfine splitting are compatible with the presence of large contributions from higher orders 
up to the highest temperature considered.\\
\indent In the temperature range explored, the origin of these terms can be traced back to higher order effects either in the
spin-dependent potential or in the spin-independent wave-function. In the latter case the true wave
function of the screening state would be more peaked at the origin, which in turn implies a stronger binding
potential with respect to the leading one of $O(g^2)$ in eq.~(\ref{eq:Schrodinger}). Since the $O(g^3)$ term in
the potential takes contribution of non-perturbative origin~\cite{Linde:1980ts}, a higher order calculation of both
the spin-independent screening masses and the hyperfine splitting has to take into account non-perturbative
contributions.\\
\indent From a more theoretical point of view, these results call for
non-perturbative computations in the three-dimensional theory in order to
match the effective theory with QCD non-perturbatively. This will allow to shed light on
the origin of the various contributions, and to verify non-perturbatively the effective theory paradigm.
At the same time, the possibility of studying QCD at high temperatures by Monte Carlo
simulations, makes the effective theory analysis less compelling especially if higher and higher
orders must be taken into account.\\
\indent Shortly before we completed the writing of this paper, a study has been
published on arXiv~\cite{Bala:2025ilf} which addresses the same topic
studied here. In particular the authors explore the effects of including
non-perturbative contributions in the spin-independent potential to
explain the hyperfine splitting observed in Ref.~\cite{DallaBrida:2021ddx}.

\section*{Acknowledgements}
We wish to thank Mikko Laine for several discussions on the topics of this paper. This work is partially
supported by ICSC – Centro Nazionale di Ricerca in High Performance Computing, Big Data and Quantum Computing,
funded by European Union – NextGenerationEU. The research of M.C. is funded through the MUR program for
young researchers “Rita Levi Montalcini”. D.L. is supported by ERC grant
StrangeScatt-101088506.

\appendix
\section{Simplification of the fermionic action \label{App:A}}
For the computation of the leading contribution to the hyperfine splitting in the mesonic screening
masses, the action in eq. \eqref{eq:NRQCD} can be further simplified and written in a more compact way.
In particular, we can safely neglect the interacting terms appearing in the transverse covariant
derivative $D_\perp$ because they would produce either spin-independent
or sub-leading spin-dependent contributions. For each flavour, the action in eq.~\eqref{eq:NRQCD} can
then be written as 
\begin{align}
  S_{\rm NRQCD} = i  \int d^3 x \, &\Bigg\{\bar\chi \left[M +\partial_3 -g_{\rmii E} {\cal A}^+ -
    \frac{1}{2\pi T}\left(\nabla_\perp^2+\frac{g_{\rmii E}}{4i} [\sigma_j,\sigma_k] F_{jk} \right)\right] \chi \nonumber\\
  +\,&\hspace{0.35cm} \bar\phi \left[ M - \partial_3 - g_{\rmii E} {\cal A}^- -\frac{1}{2\pi T} \left( \nabla_\perp^2 +
    \frac{g_{\rmii E}}{4i} [\sigma_j,\sigma_k] F_{jk} \right) \right] \phi \Bigg\}+ \dots\, ,
\end{align}
where we introduced the spin-independent vertex operator ${\cal A}^\pm$ defined as
\begin{align}
\label{eq:Apm}
    {\cal A}^\pm = {\cal A}^{\pm,\lambda} T^\lambda = \left( A^\lambda_0 \pm i A^\lambda_3 \right) T^\lambda \,  , \qquad \lambda=1,\dots,N_c^2-1\,.
\end{align}
Furthermore, by recalling that $[\sigma_j,\sigma_k]=2i\epsilon_{jkl} \sigma_l$ and since we are only interested
in $j,k=1,2$ and therefore $l=3$, the spin-dependent term in the action can be readily written as
\begin{align}
\label{eq:B}
    [\sigma_j,\sigma_k]F_{jk} = -4i\sigma_3 B_3\,,
\end{align}
where the chromo-magnetic field is defined as 
\begin{align}
\label{eq:magnetic_field}
    B_3  = -\Big[ \partial_1 A_2 -\partial_2 A_1 \Big] + O(g_{\rmii E})\, ,
\end{align}
and we can safely neglect $O(g_{\rmii E})$ terms since we are only interested in the leading
spin-dependent corrections. By introducing the differential operators
\begin{align}
\label{eq:D}
    {\cal D}^\pm = M \pm \partial_3 - \frac{\nabla_\perp^2}{2\pi T} \, ,
\end{align}
and the interaction vertices
\begin{align}
\label{eq:Vpm}
    {\cal K}^\pm (x) = {\cal K}^{\pm,\lambda} (x) T^\lambda = \left({\cal A}^{\pm,\lambda} (x)\id - \frac{1}{2\pi T} \sigma_3B_3^{\lambda}(x) \right) T^\lambda \,,
\end{align}
with $\id$ being the identity matrix in the spin indices, the NRQCD action can be written in the
compact way given in eq. \eqref{eq:NRQCD2}.

\section{Propagators in the free theory}
\label{App:B}
In this appendix we define the free propagators for the fields entering the actions
$S_\rmii{EQCD}$ and $S_{\rmii{NRQCD}}$ in eqs.~(\ref{eq:EQCD}) and (\ref{eq:NRQCD}), respectively.
By setting $g_{\rmii E}=0$ in these actions, the free-theory gauge field propagators read
\begin{align}
    \label{eq:gauge_prop}
    \expval{A_\mu^\lambda(x) A_\nu^\rho(0)}_0 = \delta^{\lambda\rho} \Delta_{\mu\nu} (x)\, ,
\end{align}
where $\mu,\nu=0,\dots,3$, $\lambda,\rho=1,\dots, N_c^2-1$ and in the Feynman gauge it holds
\begin{align}\label{eq:gauge_prop_2}
    \Delta_{\mu\nu} (x) =\delta_{\mu\nu} \int \frac{\diff^3 p}{(2\pi)^3} e^{ip\cdot x} \left[ \frac{\delta_{\mu 0}}{p^2+m_{\rmii E}^2} + \frac{\delta_{\mu i}}{p^2} \right] \, , 
\end{align}
with $m_{\rmii E}$ being defined in eq. \eqref{eq:ge}, and $i=1,2,3$.
The quark propagators at tree-level are defined as
\be
\expval{\chi(x)\bar\chi (0)}_0 = S^{(0)}_\chi(x,0)\,, \qquad
\expval{\phi (x)\bar\phi (0)}_0 = S^{(0)}_\phi(x,0) \, , 
\ee
where
\begin{align}
\label{eq:quark_prop_free_app}
    S^{(0)}_\chi (x,0) = S^{(0)}_\phi (0,x) = -i\theta (x_3)\id  \int_{\mathbf{p}} e^{i\mathbf{p}\cdot \mathbf{r}} e^{-x_3 \left( M+\frac{\mathbf{p}^2}{2\pi T} \right)}\, ,\qquad x=(\mathbf{r},x_3)\,, 
\end{align}
and $\id$ is the identity in spinor and colour indices. The integration over the transverse momenta
is defined as $\int_{\mathbf{p}} = \int\diff^2 \mathbf{p}/(2\pi)^2$.

\section{Equations of motion at next-to-leading order}
\label{App:C}
In this appendix we report the intermediate steps to obtain eq.~\eqref{eq:EOM-next-to-leading-1} from 
eq.~\eqref{eq:EOM-next-to-leading}. By inserting eq.~(\ref{eq:quark_prop_nlo_1}) into the r.h.s. of eq.
\eqref{eq:EOM-next-to-leading}, we obtain
\begin{align}
  &\Biggl[ 2M + \partial_3 -\sum_{i=1,2} \frac{\nabla_{\mathbf{r}_i}^2}{2\pi T} \Biggr] \expval{W_{\cal O}(\mathbf{r}_1,\mathbf{r}_2;x_3)} =
  g_{\rmii E}^2\, \Bigg\langle
  \Tr \Big\{ \Big[{\cal A}^+(\mathbf{r}_1,x_3) +{\cal A}^-(\mathbf{r}_2,x_3)\Big] \\ 
 & \Big[\Sigma_{\cal O}\, {\cal S}^{(0)}_+(\mathbf{r}_1,x_3)\, \Sigma_{\cal O}\, {\cal S}^{(1)}_- (\mathbf{r}_2,x_3) +
    \Sigma_{\cal O}\, {\cal S}^{(1)}_+ (\mathbf{r}_1,x_3)\, \Sigma_{\cal O}\, {\cal S}^{(0)}_- (\mathbf{r}_2,x_3)  \Big]\Big\}
   -\!\! \frac{1}{2\pi T}\!\! \Tr \biggr\{\!\Big[s_{\cal O} B_3(\mathbf{r}_1,x_3)\nonumber\\
  & +\!\! B_3(\mathbf{r}_2,x_3)\Big] \sigma_3 \Big[
           \Sigma_{\cal O} {\cal S}^{(0)}_+(\mathbf{r}_1,x_3) \Sigma_{\cal O} {\cal S}^{(1)}_-(\mathbf{r}_2,x_3)
     \!+\! \Sigma_{\cal O} {\cal S}^{(1)}_+(\mathbf{r}_1,x_3) \Sigma_{\cal O} {\cal S}^{(0)}_-(\mathbf{r}_2,x_3)
    \Big]\!\biggr\}\!\!\Bigg \rangle + O(g^3_{\rmii E})\, . \nonumber
\end{align}
By inserting the expression for the quark propagators at next-to-leading order in eq. \eqref{eq:quark_prop_nlo} and
by recalling that, given the expression for ${\cal K}^{\pm}$ in eq. \eqref{eq:Vpm}, ${\cal A}^\pm$ only contracts with ${\cal A}^\pm$
and $B_3$ only with $B_3$, the equation of motion can be written as 
\begin{align}
\begin{aligned}
  \Biggl[& 2M + \partial_3 -\sum_{i=1,2} \frac{\nabla_{\mathbf{r}_i}^2}{2\pi T} \Biggr]
  \expval{W_{\cal O}(\mathbf{r}_1,\mathbf{r}_2;x_3)}
  = \frac{g_{\rmii E}^2}{N_c}\,  W^{(0)}_{\cal O}(\mathbf{r}_1,\mathbf{r}_2;x_3) \times \\
  &  \biggl\{ \expval{\Tr\bigg\{\Big[{\cal A}^+(\mathbf{r}_1,x_3) + {\cal A}^-(\mathbf{r}_2,x_3) \Big]
    \int_0^{x_3} \diff z_3 \Big[ {{\cal A}^+(\frac{z_3}{x_3}\mathbf{r}_1,z_3) +
                                \cal A}^-(\frac{z_3}{x_3} \mathbf{r}_2,z_3)\Big]\bigg\}} + \\
  & \frac{1}{(2\pi T)^2} \int_0^{x_3} \diff z_3\, \Big\langle \Tr\bigg\{B_3(\mathbf{r}_1,x_3)
  B_3(\frac{z_3}{x_3} \mathbf{r}_1,z_3)\, +
  B_3(\mathbf{r}_2,x_3) B_3(\frac{z_3}{x_3}\mathbf{r}_2,z_3)\bigg\} \Big\rangle\, + \\
  & \frac{s_{\cal O}}{(2\pi T)^2} \int_0^{x_3} \diff z_3\, \Big\langle \Tr\bigg\{
  B_3(\mathbf{r}_1,x_3) B_3(\frac{z_3}{x_3} \mathbf{r}_2,z_3) +
  B_3(\mathbf{r}_2,x_3) B_3(\frac{z_3}{x_3} \mathbf{r}_1,z_3) \bigg\}
  \Big\rangle \biggr \} + O(g^3_{\rmii E}) \,, \\
\end{aligned}
\end{align}
where the tree-level value $W^{(0)}_{\cal O}(\mathbf{r}_1,\mathbf{r}_2;x_3)$ is defined in eq.~\eqref{eq:W0},
and we have introduced $s_{\cal O}=(+1,-1)$ for the pseudoscalar and the vector fields respectively.
Moreover we have exploited the fact that, when contracting the gauge fields
$\expval{{\cal A}^{\pm,\rho} {\cal A}^{\pm,\lambda}} \propto \delta^{\rho\lambda}$ and
$\expval{B_3^\rho B_3^\lambda} \propto \delta^{\rho\lambda}$.

\section{Evaluation of the static potential}
\label{App:D}
In this appendix we explicitly show the intermediate steps which lead to the final expression of the static potentials in eq. \eqref{eq:static_SI} and \eqref{eq:static_SD}.

\subsection{Spin-independent contribution}
Starting from eq. \eqref{eq:USI1}, the leading spin-independent static potential can be written as
\begin{align}
& {\cal U}_{{\rm SI}_1}(\mathbf{r}_1,\mathbf{r}_2;x_3) = \!  -\!\frac{g_{\rmii E}^2}{N_c} \!
    \Big\langle\Tr\Big\{\!\Big[ {\cal A}^+(\mathbf{r}_1,x_3)\! +\! {\cal A}^-(\mathbf{r}_2,x_3)\Big]\times\\[0.25cm]
    &\hspace{-0.25cm} \int_0^{x_3}\!\!\!\!\! \diff z_3
    \Big[ {\cal A}^+(\frac{z_3}{x_3}\mathbf{r}_1,z_3)\! +\! {\cal A}^-(\frac{z_3}{x_3}\mathbf{r}_2,z_3)\Big]\Big\}\!\Big\rangle\!
    = 2 g_{\rmii E}^2 C_F \Big[ {\cal V}^- \left(\mathbf{r}_1,\mathbf{r}_2,x_3 \right) - {\cal V}^+ \left( \mathbf{r}_1,\mathbf{r}_1,x_3 \right) \Big] \, .\nonumber
\end{align}
where, we contracted the gauge fields and, similarly to the procedure used in Ref. \cite{Giusti:2024mwq}, we introduced the integrals
\begin{align}
  {\cal V}^{\pm} (\mathbf{r}_1, \mathbf{r}_2,x_3) = -\int_0^{x_3} \diff z_3
  \left[ \Delta_{33} \left( \mathbf{r}_1-\mathbf{r}_2 + \frac{z_3}{x_3}\mathbf{r}_2 ,z_3\right) \mp
    \Delta_{00}\left( \mathbf{r}_1-\mathbf{r}_2 \frac{z_3}{x_3}\mathbf{r}_2,z_3\right) \right] \, ,
\end{align}
where $\Delta_{\mu\nu}$ is the gauge field propagator defined in eq. \eqref{eq:gauge_prop}, in which we made explicit the dependence on the transverse and longitudinal coordinates. Notice that it holds ${\cal V}^\pm (\mathbf{r}_1,\mathbf{r}_2,x_3)={\cal V}^\pm (\mathbf{r}_2,\mathbf{r}_1,x_3)$ under the exchanges $\mathbf{r}_1\leftrightarrow \mathbf{r}_2$.
By taking the large separation limit in the third spatial direction, see eq. (C.7) and (C.9) of Ref. \cite{Giusti:2024mwq},
we arrive at
\begin{align}
  U_{\rm {SI}_1}\left( \mathbf{r}_1 -\mathbf{r}_2 \right) = \lim_{x_3\to \infty} {\cal U}_{\rm {SI}_1} \left(\mathbf{r}_1,\mathbf{r}_2,x_3 \right) = \frac{g_{\rmii E}^2 C_F}{2\pi} \bigg[ \ln \left(\frac{m_{\rmii E}r}{2}\right) +
    \gamma_{\rmii E} - K_0(m_{\rmii E}r) \bigg] \, ,
\end{align}
where we introduced $r=|\mathbf{r}_1-\mathbf{r}_2|$, $\gamma_{\rmii E}$ is the Euler-Mascheroni constant, $K_0$ is a modified Bessel function, and $m_{\rmii E}$ is the Debye mass introduced in eq. (\ref{eq:ge}). The potential above is exactly the expression
reported in \eqref{eq:static_SI}, and which was found in Ref. \cite{Laine:2003bd}.

\subsection{Spin-dependent contribution}
Analogously, the temperature-suppressed, spin-dependent potential in eq. \eqref{eq:US} can be written as
\begin{align}
\label{eq:calUS}
    {\cal U}_{\cal O} (\mathbf{r}_1, \mathbf{r}_2, x_3) = - g_{\rmii E}^2 \frac{2 s_{\cal O} C_F}{(2\pi T)^2} {\cal B}(\mathbf{r}_1,\mathbf{r}_2,x_3) \, ,
\end{align}
where we introduced the definition
\be
\hspace{-0.25cm}    {\cal B}(\mathbf{r}_1,\mathbf{r}_2,x_3)\!\! =\!\! \frac{1}{N_c^2-1}\!\! \int_0^{x_3}\!\!\!\!\! \diff z_3
    \Big\langle \Tr\Big\{ B_3(\mathbf{r}_1,x_3)B_3\Big(\frac{z_3}{x_3}\mathbf{r}_2,z_3\Big)\!\!+\!\!
    B_3(\mathbf{r}_2,x_3)B_3\Big(\frac{z_3}{x_3}\mathbf{r}_1,z_3\Big)\Big\} \Big\rangle .
\ee
By taking the contractions of the gauge fields in eq. \eqref{eq:gauge_prop}, and by expressing the chromo-magnetic field
as in eq. \eqref{eq:magnetic_field}, the integral can be written as\footnote{Here the superscript in the partial derivative refers to which $\mathbf{r}_i$ (either $\mathbf{r}_1$ or $\mathbf{r}_2$) the derivative is taken with respect to. Therefore $\partial_i^{\mathbf{r}_j}$ indicates the partial derivative with respect to the $i$th component of the $\mathbf{r}_j$ coordinate.}
\bea
\label{eq:calB}
B\left( \mathbf{r}_1 -\mathbf{r}_2 \right) &= & \lim_{x_3\to \infty} {\cal B}\left( \mathbf{r}_1, \mathbf{r}_2,x_3 \right)\\
& = &
\lim_{x_3\to \infty} \int_0^{x_3} \diff z_3 \Big[ \partial_1^{\mathbf{r}_2} \partial_1^{\mathbf{r_1}} \Delta_{22} (\mathbf{r}_1-\mathbf{r}_2,x_3-z_3) + \partial_2^{\mathbf{r}_2} \partial_2^{\mathbf{r_1}} \Delta_{11} (\mathbf{r}_1-\mathbf{r}_2,x_3-z_3) \Big]  \, .\nonumber
\eea
By recalling the expression of the spatial components of the gluon propagators, see eq. \eqref{eq:gauge_prop_2},
the derivative terms in the expression above read
\begin{align}
  \partial_j^{\mathbf{r}_2} \partial_j^{\mathbf{r}_1} \Delta_{ii} (\mathbf{r}_1-\mathbf{r}_2,x_3-z_3) =
  \int\frac{\diff^3 p}{(2\pi)^3} e^{i\mathbf{p}\left(\mathbf{r}_1-\mathbf{r}_2\right)} e^{i p_3 (x_3-z_3)} \, \frac{p_j^2}{p^2} \, .
\end{align}
By inserting this expression in eq. \eqref{eq:calB}, the integral can be written as
\begin{align}
  B\left(\mathbf{r}_1 - \mathbf{r}_2\right) = \lim_{x_3\to \infty}
  \int_0^{x_3} \diff z_3 \int \frac{\diff^3 p}{(2\pi)^3} e^{i\mathbf{p}\left(\mathbf{r}_1-\mathbf{r}_2\right)} e^{i p_3(x_3-z_3)} \left[ 1-\frac{p_3^2}{p^2} \right] \,,
\end{align}
where we used the fact that $p^2-p_3^2=p_1^2+p_2^2 = \mathbf{p}^2$. Therefore the integral above can be split into two different contributions as
\bea
B \left( \mathbf{r}_1 - \mathbf{r}_2\right) & = & \lim_{x_3\to \infty} \Big[\delta^{(2)}(\mathbf{r}_1-\mathbf{r}_2) \int_0^{x_3} \diff z_3 \,\delta(x_3-z_3) - {\cal R} \left( \mathbf{r}_1 - \mathbf{r}_2,x_3 \right) \Big] \\[0.25cm]
& = & \frac{1}{2} \delta^{(2)}(\mathbf{r}_1-\mathbf{r}_2) - \lim_{x_3\to \infty}
{\cal R} \left( \mathbf{r}_1 - \mathbf{r}_2,x_3 \right)\, ,\nonumber
\eea
where we have introduced the remainder time-integral
\begin{align}
\label{eq:calR}
    {\cal R} \left( \mathbf{r}_1- \mathbf{r}_2,x_3 \right) =  \int \frac{\diff p_3}{2\pi}\, p_3^2 \int_0^{x_3} \diff z_3 \, e^{i p_3(x_3-z_3)} \int \frac{\diff^2 \mathbf{p}}{(2\pi)^2} \, \frac{ e^{i\mathbf{p}\left(\mathbf{r}_1-\mathbf{r}_2\right)}}{\mathbf{p}^2+p_3^2} \, .
\end{align}
The integration over the transverse components of the momentum can be carried out by recalling that it holds
\begin{align}
    \int \frac{\diff^2 \mathbf{p}}{(2\pi)^2} \, \frac{ e^{i\mathbf{p}\left(\mathbf{r}_1-\mathbf{r}_2\right)}}{\mathbf{p}^2+p_3^2} = \frac{1}{2\pi} K_0\big(|p_3|\, r\big) \,,
\end{align}
where $K_0$ is a modified Bessel function. Analogously by performing the integration over the longitudinal spatial coordinate, the
integral in eq. \eqref{eq:calR} is reduced to a one-dimensional integral over the $p_3$ variable that reads
\begin{align}
    {\cal R} \left( \mathbf{r}_1 - \mathbf{r}_2,x_3 \right) = \frac{i}{2\pi} \int \frac{\diff p_3}{2\pi} \,p_3\left( 1-e^{ip_3 x_3} \right) K_0\big(|p_3|\, r\big).
\end{align}
Finally the integral over the longitudinal momentum can be carried out by taking into account the symmetry
properties of the integrand under the change  $p_3\to -p_3$, yielding to
\begin{align}
\label{eq:calR_final}
      {\cal R} \left( \mathbf{r}_1 - \mathbf{r}_2,x_3 \right) = \frac{1}{2\pi^2}\int_0^{\infty}
      \diff p_3 \, p_3 \sin\left( p_3 x_3 \right) K_0\left(p_3\, r \right)=
      \frac{x_3}{4\pi\left( x_3^2 +r^2 \right)^{3/2}} \,.
\end{align}
Therefore by taking the large separation limit in the third spatial direction in eq. \eqref{eq:calB}, and by recalling that,
according to eq. \eqref{eq:calR_final}, $\displaystyle \lim_{x_3\to\infty}{\cal R}\left( \mathbf{r}_1 - \mathbf{r}_2,x_3 \right) =0$, we obtain
\begin{align}
    B\left(\mathbf{r}_1 - \mathbf{r}_2\right) = \frac{1}{2} \delta^{(2)}(\mathbf{r}_1-\mathbf{r}_2) \,.
\end{align}
Finally, by inserting this expression into eq. \eqref{eq:calUS} we obtain the final expression for the spin-dependent static potential
\begin{align}
    U_{\cal O} (\mathbf{r}_1 - \mathbf{r}_2) = \lim_{x_3\to \infty} {\cal U}_{\cal O} \left(\mathbf{r}_1,\mathbf{r}_2,x_3 \right) = -g_{\rmii E}^2 \frac{s_{\cal O} C_F}{(2\pi T)^2} \delta^{(2)}(\mathbf{r}_1-\mathbf{r}_2) \,
\end{align}
reported in eq. \eqref{eq:static_SD}.

%\{bibliographystyle}{unsrt}
\bibliographystyle{JHEP}
%\addcontentsline{toc}{section}{Bibliography}
\bibliography{biblio}

\end{document}